\documentclass[twocolumn]{aa}
\usepackage{graphicx}
\usepackage{txfonts}
\begin{document}
   \title{The baryon mass function for galaxy clusters}
\titlerunning{Clusters baryon mass function}

   \author{S.~E.~Nuza
\inst{1}
          \and
A.~Blanchard
\inst{2}
  }

   \offprints{sebasn@iafe.uba.ar}

   \institute{Instituto de Astronom\'{\i}a y F\'{\i}sica del Espacio, C.C. 67, 
Suc. 28, 1428, Buenos Aires, Argentina\\
 \email{sebasn@iafe.uba.ar}
\and
Laboratoire d'Astrophysique de Toulouse--Tarbes,
14 Avenue E.~Belin, F--31400 Toulouse, France\\
                    \email{alain.blanchard@ast.obs-mip.fr}
             }
\date{}

\abstract{The evolution of the cluster abundance with redshift is known
to be a powerful cosmological constraint when applied to X-ray clusters.
Recently, the evolution of the baryon mass function has
been proposed as a new variant that is free of the uncertainties
present in the temperature-mass relation. A flat model with
$\Omega_{\rm M} \simeq 0.3$ was shown to be preferred in the case of
a standard cold dark matter scenario.}
{We compared the high redshift predictions of the baryon mass in clusters 
with data for a more general class of spectra with a varying $shape$ factor 
$\Gamma$ without any restriction to the standard cold dark matter scenario in 
models normalized to reproduce the local baryon mass function.} 
{Using various halo mass functions existing in the literature we 
evaluated the corresponding baryon mass functions for the case of the 
non-standard power spectra mentioned previously.}
{We found that models with $\Omega_{\rm M} \simeq 1$ and 
$\Gamma \simeq 0.12$ reproduce high redshift cluster data just as well 
as the concordance model does.}
{Finally, we conclude that the baryon mass function evolution alone does 
not efficiently discriminate between the more general family of flat 
cosmological models with non-standard power spectra.}


   \maketitle
%

\section{Introduction}

The growth of structure in the Universe is believed
to be the result of gravitational collapse generated
by the existence of tiny departures from homogeneity and
isotropy (presumably generated during inflation) in the
primordial distribution of matter (see e.g., Peacock 1999).
The overdensity of matter in a particular comoving scale $l$ 
evolves according to linear theory until it reaches
a value of $\delta_k \sim 1$, where we have used $k \equiv |\bf{k}|$
with $\bf{k}$ the comoving {\it wavevector} that satisfies the 
relation $|\bf{k}|$=$2\pi/l$.
For later times the evolution is highly nonlinear and the
formation of bound structures like galaxies and  galaxy clusters 
takes place. It is believed that this hierarchical process of 
structure formation is still at work today.
Press \& Schechter (1974, hereafter PS) developed a semianalytical
formulation to deal with this regime and eventually predict
the number of collapsed objects (often called {\it virialized}
objects) of a given mass $M$ (associated with the scale $l$)
at a given redshift $z$, i.e. to determine the so-called {\it mass function}.
Interest in the PS approach has grown in recent years because it
appears to reproduce the results of  numerical simulations well (e.g. 
Efstathiou et al. 1988; White, Efstathiou \& Frenk 1993; Lacey \& Coley 1994) 
where the nonlinear regime can be tested unambiguously inside the limits 
imposed by the resolution of the simulation.
Sheth \& Tormen (1999) obtained a different expression for the mass function 
by assuming an {\it ellipsoidal} collapse instead of a {\it spherical} one 
(as assumed in the PS theory) and found better agreement between the 
model and the numerical results. Sheth, Mo, \& Tormen (2001, hereafter SMT) 
present an improved version of their 1999 work. Instead of working out a 
semianalytical approach Jenkins et al. (2001, hereafter J01) have 
introduced several fits to the numerical simulations using different 
algorithms for the halo finder and for several kinds of cosmologies. 
Further analyses have been made of these matters such as, White (2002) 
and Warren et al. (2005).
Of particular interest is the evolution of the mass function with 
redshift that has been shown to be sensitive primarily to the mass density 
of the Universe (Blanchard \& Bartlett 1998). This high sensitivity allows 
us to use the evolution of the abundance of X-ray clusters as a powerful 
cosmological test (Oukbir \& Blanchard 1992). 
Recently, the {\it baryon} mass function (i.e., the number density of 
comoving objects with a given baryon mass) has been advocated by Vikhlinin 
et al. (2003, hereafter V2003) as an useful alternative and cosmological 
constraints were derived in the context of standard cold dark matter 
(hereafter CDM) spectrum.
V2003 have consequently concluded that cluster data favors a concordance-like 
Universe. This analysis seems to conflict with the study made by 
Blanchard et al. (2000) and with the  recent 
$XMM-Newton~\Omega$-project (Vauclair et al. 2003),
which is somewhat surprising because Sadat et al. (2005) found that gas 
fraction in distant clusters within the $XMM-Newton~\Omega$-project was 
consistent with a high-density Universe and not with a concordance model.
These two sets of analyses therefore suggest that the baryon mass fraction
should also  be consistent with a high-density Universe. 
The present paper aims to clarify this issue. Here we study the more 
general class of spectra with a varying {\it shape} factor $\Gamma$, 
without any restriction to the standard CDM model.
In Sect. 2 we briefly review the PS formalism and recall the 
results of SMT and J01. In Sect. 3 we introduce the 
temperature-mass (hereafter $T-M$) relation, and we use the different 
available expressions for the baryon mass function, normalized to local 
cluster data, to compare them with the high redshift observations provided 
by V2003. 
Finally, we discuss our main conclusion that data on the baryon mass function 
of clusters can also be reproduced in a critical Universe with 
$\Omega_{\Lambda} \simeq 0$, indicating that there is no discrepancy in 
both approaches.

\section{Mass functions of galaxy clusters}

\subsection{The PS formalism}
 
The PS approach is based on the assumption of an initial {\it Gaussian} 
overdensity field $\delta(\mathbf{x},z_{\rm i})$ and a {\it spherical} 
model for the subsequent collapse (Partridge \& Peebles 1967). 
Let $n(M,z)$ be the comoving number density of objects with mass $M$ at 
a given redshift $z$. Then,

\begin{equation}
n(>M,z)=\int_{M}^{+\infty} \frac{dn(M',z)}{dM'}dM'
\end{equation}

\noindent is the number of collapsed objects of mass greater than $M$ 
at redshift $z$. The {\it mass function} resulting from these priors 
reads as (for a detailed discussion see e.g. 
Blanchard, Valls-Gabaud \& Mamon 1992)

\begin{equation}
\label{PS}
\frac{dn(M,z)}{dM}= \frac{\rho_0}{M}
\frac{d\nu}{dM}{\cal F}(\nu).
\end{equation}

\noindent In this formula, $\rho_0$ is the comoving background density 
of the Universe, $\nu \equiv \delta_{\rm c}/\sigma(M,z)$, where 
$\delta_{\rm c}$ is the {\it linear} overdensity evaluated at the 
virialization time, $\sigma(M,z)$ is the {\it r.m.s.} amplitude of the 
matter fluctuations at a given mass scale $M$, and $\cal F$$(\nu)$ is a 
function taken as 

\begin{equation}
{\cal F}(\nu)=
\sqrt{\frac{2}{\pi}}
\exp\left(-\frac{\nu^{2}}{2}\right)
\end{equation}

\noindent in the original PS work. 
For an Einstein-de Sitter Universe, the value for $\delta_{\rm c}$ 
is approximately 1.69. This value is normally assumed because of 
the weak cosmological dependence of the linear overdensity at 
virialization (e.g. Colafrancesco \& Vittorio 1994).
In order to evaluate $\sigma(M,z)$ one has to smooth the density
field $\delta(\mathbf{x},z)$ with some known {\it window function}
$W_k$ for a given $k$. The expression for $\sigma(M,z)$ results in

\begin{equation}
\label{sigma}
\sigma^2(M,z)= \int_0^{\infty} \frac{dk}{2\pi^2}k^2 |\delta_k(z)|^2 F^2 [W_k],
\end{equation}

\noindent where $F[W_k]$ is the Fourier transform of $W_k$, and 
$|\delta_k(z)|^2$ is the power spectrum of $\delta(\mathbf{x},z)$. 
The most popular election for $W_k$ is a spherical top-hat in real 
space such that the relation between the mass scale $M$ and the comoving 
linear scale $l$ is given by $l^3=6M/\pi \rho_0$. 
From Eqs. (\ref{PS}) and (\ref{sigma}), it is easy to see that all 
cosmological dependence enters through the evolution of the {\it linear} 
overdensity field of matter, so by setting its value adequately one can apply 
this formalism to any cosmology of interest.

\subsection{Improvements to PS theory}

\begin{table}
\begin{center}
\caption{Different fits for $f(\sigma,z)$ provided by
J01  for different cosmological models and various halo finders. 
CM and HF stands for Cosmological Model and Halo Finder respectively.
$fof(a)$ refers to a {\it friend of friend} algorithm with 
an interparticle separation $a$ and $so(\Delta)$ refers to a
{\it spherical overdensity} algorithm with contrast density $\Delta$ 
(respect to the background). $All$ means all cosmological models 
listed in table 2 of J01 paper.}
\vspace{2pt}
\begin{tabular}{ccccccr} \hline  
\# & A & B & C & CM & HF & $\Delta$ \\
\hline
1 & 0.307 & 0.61 & 3.82 & $\tau$CDM & $fof(0.2)$ & $\simeq$ 180\\
2 & 0.301 & 0.64 & 3.88 & $\Lambda$CDM & $fof(0.164)$ & $\simeq$ 324\\
3 & 0.301 & 0.64 & 3.82 & $\tau$CDM & $so(180)$ & 180\\
4 & 0.316 & 0.67 & 3.82 & $\Lambda$CDM & $so(324)$ & 324\\
5 & 0.315 & 0.61 & 3.80 & $All$ & $fof(0.2)$ & $\simeq$ 180\\
\hline
\end{tabular}
\end{center}
\label{Jfits}
\end{table}

As mentioned in the introduction, SMT have introduced an improved
version for the mass function of collapsed objects. Their approach
is similar to that of PS, but instead of assuming a spherical model
for virialization, they used elliptical collapse and obtained 
a somewhat different expression that agrees better with $N$-body 
numerical simulations. The SMT expression reads as follows

\begin{equation}
\label{SMT}
\frac{dn(M,z)}{dM}=c\sqrt{\frac{2a}{\pi}}\frac{\rho_0}{M}
\frac{d\nu}{dM}
\left(1+\frac{1}{(a\nu^2)^p}\right)
\exp\left(-\frac{a\nu^2}{2}\right),
\end{equation}

\noindent with $a=0.707$, $c=0.3222$, and $p=0.3$. Setting $a=c=1$ and 
$p=0$ in this formula leads to the PS formalism.
More recently, J01 have found several fits to the mass function using 
the results of their $N$-body numerical simulations. 
Specifically, they consider the quantity

\begin{equation}
\label{f}
f(\sigma,z) \equiv \frac{M}{\rho_0}\frac{dn}{d\ln \sigma^{-1}}
\end{equation}

\noindent that is parametrized assuming the following functional form
motivated by the {\it ansatz} given by Eqs. (\ref{PS}) 
and (\ref{SMT}):

\begin{equation}
\label{f_exp}
f(\sigma,z) \equiv A \exp \left(-| \ln \sigma^{-1} + B |^C\right),
\end{equation}

\noindent with $A$, $B$, and $C$ the fitting parameters.
In particular they use two distinct ways of object grouping,
namely {\it friend-of-friend} and {\it spherical-overdensity}
halo finders, for various types of cosmologies (see their paper for
details). In Table 1 we give the different values of the
fitting parameters in these several situations. An important quantity
related of these halo finders is the cluster density with 
respect to the background Universe density (or {\it contrast} density). 
In general, its value depends on the cosmology and redshift. 
J01 assume a {\it constant} value for the contrast 
density (see $\Delta$ in table 1) in order to get their fits.

\section{Determination of $\Omega_{\rm M}$}

\subsection{The Temperature-Mass relation}

An unavoidable ingredient in the determination of the mass function
for galaxy clusters was until recently the use of the $T-M$ relation, 
i.e. the relation between the cluster total (virial) mass and the (observed) 
X-ray temperature. It allows determination of the mass function so that its 
evolution can then be used to constrain the value of $\Omega_{\rm M}$ 
(e.g. Oukbir \& Blanchard 1992, 1997). Standard scaling laws 
(e.g. Kaiser 1986) allow us to write the $T-M$ relation as follows

\begin{equation}
\label{T-M}
T_{\rm X} = A_{TM} \left(\Omega_{\rm M}
\frac{\Delta(\Omega_{\rm M},z)}{178}\right)^{1/3} M_{15}^{2/3}h^{2/3}(1+z),
\end{equation}

\noindent where $\Delta(\Omega_{\rm M},z)$ is the contrast density
mentioned in 2.2 and $h$ the present Hubble constant in units of 
100 Km s$^{-1}$ Mpc$^{-1}$. The subscript 15 means that masses are taken 
in units of $10^{15} M_{\sun}$. $ A_{TM} $ is a normalization factor 
in that case.
A known uncertainty exists in $A_{TM}$ because the value that results
from numerical simulations is significantly different from the one based on 
the hydrostatic equation (Roussel et al. 2000). Despite this, the $T-M$ 
relation has been applied in the past to link cluster observations 
($\propto T_{\rm X}$) with cluster mass.
A conservative approach to this was assumed by Vauclair et al. (2003) when 
using two extreme normalizations in their analysis of the 
$XMM-Newton~\Omega$-project (Lumb et al. 2004). In particular, their 
conclusions are roughly independent of the $A_{TM}$ value. However, Blanchard 
and Douspis (2005) have recently introduced a procedure to remove this 
uncertainty.
In the next subsection we review the recent proposal by V2003 to avoid 
the $T-M$ normalization problem applied to the constraint of $\Omega_{\rm M}$.

\subsection{The baryon mass function}

This method relies on the standard assumption that the baryon fraction
within the virial radius in clusters should be close to the average
value in the Universe, i.e. $f_{\rm b} \simeq \Omega_{\rm b}/\Omega_{\rm M}$ 
(White et al. 1993).
Using the baryon mass at a radius of constant baryon contrast density
(e.g. Vikhlinin, Forman \& Jones 1999) and the fact that, to a first 
order, the baryon and total mass in clusters are trivially related by
$M_{\rm b}=Mf_{\rm b}$, one can deduce the functional form
$N_{\rm b}(M_{\rm b})$ of the baryon mass function as

\begin{equation}
\label{Fb}
N_{\rm b}(>M_{\rm b}) = N(>M_{\rm b}f^{-1}_{\rm b}),
\end{equation}

\noindent where $N(M)$ is the total mass function. Eqn. (\ref{Fb}) 
allows us to evaluate the baryon mass functions for different models 
according to the various expressions presented in Sect. 2. 
By studying the evolution of $N_{\rm b}(M_{\rm b})$ for high-$z$ cluster 
data, we can constrain the density parameter previous adjustment of the mass 
function to the local cluster observations. 
In this last procedure, the shape factor $\Gamma$ of the power 
spectrum and the $\sigma_8$ parameter (formula (\ref{sigma}) evaluated 
on a scale of 8 $h^{-1}$ Mpc) can be fixed. It is worth noting that the 
baryon fraction universality implies, for a cluster, that the total matter 
density contrast equals the baryonic matter contrast, i.e. 
$\Delta=\Delta_{\rm b}$, where $b$ stands for {\it baryonic}. 
The most obvious advantage of this method is that the baryonic mass is a 
quantity that can be measured directly. We refer the reader to V2003 
for details.

\begin{figure}
\centering
\includegraphics{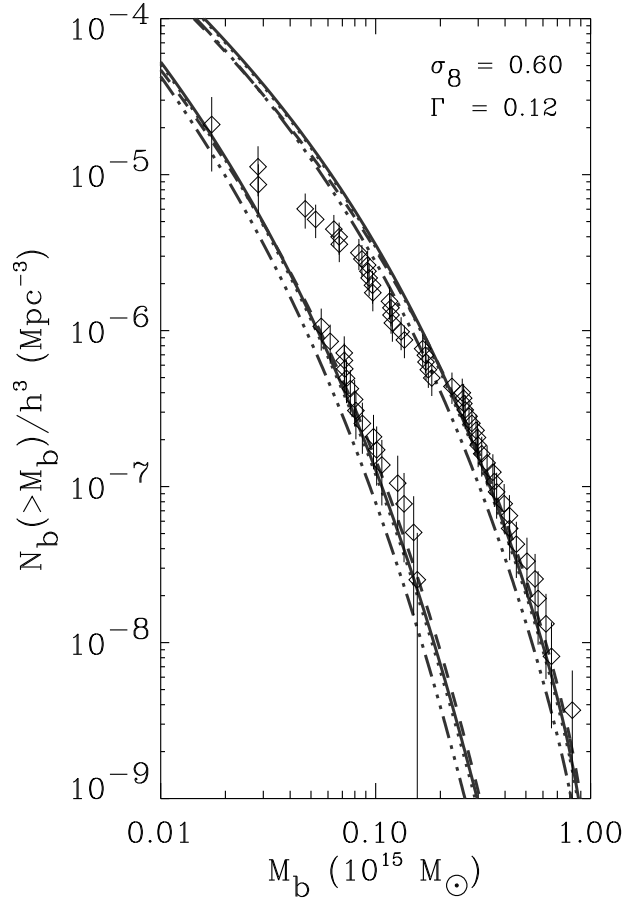}
\caption{Low and high redshift cluster data ($z \simeq 0.05$ and 
$z \simeq 0.5$ respectively, $h = 0.5$) compared with various theoretical 
baryon mass functions [dotted line: J01 \# 1, 
dashed 3-dotted line: J01 \# 3, solid line: J01 \# 5, 
dashed line: SMT, see Sect. 2]. A Universe satisfying 
($\Omega_{\rm M}$,$\Omega_{\Lambda}$) $\simeq$ (1,0) clearly fits the data 
for non-standard power spectra with $\Gamma = 0.12$.}
\end{figure}

\subsection{Standard CDM vs. $\Gamma$-varying spectra}

\begin{figure}
\centering
\includegraphics{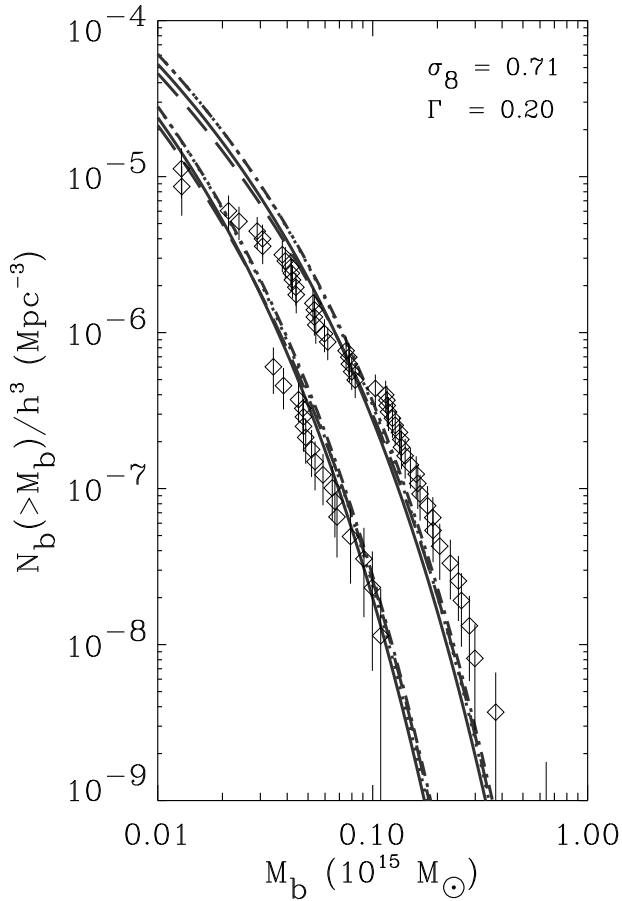}
\caption{Same as Fig. 1 but for the case of a {\it fiducial} concordance
cosmology with $h = 0.71$ [dotted line: J01 \# 2, 
dashed dotted line: J01 \# 4, solid line: J01 \# 5, 
dashed line: SMT, see Sect. 2].}       
\end{figure}

In order to make the comparisons mentioned above, we used the local
($z \simeq 0.05$) and high-$z$ ($\simeq 0.5$) cluster observations from 
Voevodkin \& Vikhlinin (2004, see their Table 1) and V2003 respectively,
where data is defined for $\Delta_{\rm b}=324$. 
When comparing the reliability of a particular cosmology to cluster 
observations, the value of $h$ can be taken arbitrarily, because once 
the baryonic mass data is scaled properly ($M_{\rm b} \propto h^{-2.25}$, 
Voevodkin \& Vikhlinin 2004), the only factor that determines 
the fit inside the context of a particular cosmology, is the redshift 
evolution of the mass function (previous constraint of the $\Gamma$ and 
$\sigma_8$ parameters using local cluster data, see Sect. 3.2). 
In our comparison we adopted two fiducial models: a concordance model 
with $\Omega_{\rm M} = 0.3$, $\Gamma = 0.2$, $\sigma_8 = 0.71$, 
$\Omega_{\rm b} = 0.04$, $h=0.71$ and a model with  $\Omega_{\rm M} = 1$, 
$\Gamma = 0.12$, $\sigma_8 = 0.6$, $\Omega_{\rm b} =0.105$, $h=0.5$. 
The first model is close to the preferred one according to V2003 when fitted 
to the baryon mass function and its evolution. The second one is the best fit 
model to X-ray cluster data proposed by Vauclair et al. (2003). 
The baryon fractions were taken to be the apparent value after computing 
the corrections of depletion and clumping (Sadat \& Blanchard 2001) using 
the analysis of local clusters by Sadat et al. 2005, i.e. 
$f_{\rm b} = \Upsilon \times \Omega_{\rm b}/ \Omega_{\rm M} \simeq 
1.14 \times \Omega_{\rm b}/ \Omega_{\rm M}$. 
When using a particular mass function that assumes another value for 
$\Delta_{\rm b}$ a correction must be applied because of the distinct 
cluster mass definition. To achieve this, a Navarro-Frenk-White 
(Navarro, Frenk \& White 1996, NFW96) universal profile for the structure of 
the CDM halos in clusters was used assuming $C = 5$ (where $C$ is the 
concentration parameter defined in Eqn. (3) of NFW96), giving 
typical corrections for the baryonic mass on the order of 20 \%. 
The results of our comparisons can be seen in Figures 1 and 2 where we have 
plotted various of the expressions for the baryon mass function 
(see Sect. 2) in each graph. As can be seen in the figures, the different 
functions lead to nearly identical behaviour for the same values of 
$\Gamma$ and $\sigma_8$ up to $z \simeq 0.5$. 
To get total agreement, only a tiny change in these parameters is needed 
for each expression. We checked that in such a case the various
expressions lead to a very similar level of evolution, since almost identical. 
We also noticed that V2003 have used a $\Upsilon$ varying parameter with X-ray 
temperature that leads to better agreement in the predicted mass function 
at low masses. The validity of such a variation with X-ray temperature is 
questionable (Sadat et al. 2005) and does not modify our conclusions. 
The most noticeable result obtained in the comparison is that the 
evolution of the baryon mass function in an Einstein-de Sitter cosmology 
($\Omega_{\Lambda} \simeq 0$) is clearly consistent with the high redshift 
cluster data. Although the amount of evolution is clearly not the same between 
different cosmologies the baryon masses inferred from high redshift data 
also differ in a way that accidentally compensates for the evolution abundance 
effect.

\section{Discussion}

We have found that a non-standard ($\Gamma$-varying) spectra can reproduce 
the observed baryon mass distribution function both for local and high 
redshifts in the case of an Einstein-de Sitter cosmology just as well as 
the concordance model does. 
For this, we used several formulas for the baryon mass function and 
found that the results are essentially insensitive to the expression used.
It is important to recall that the V2003 analysis is based on the assumption 
of standard CDM power spectra for matter and on a particular value of the 
Hubble constant. This assumption fixes the value of the shape factor 
essentially in $\Gamma \simeq \Omega_{\rm M} h$ (e.g. Peacock \& Dodds 1994). 
Furthermore, as they have used $h \simeq 0.65$, i.e. a Hubble constant value 
near the Hubble Key Project result (Freedman et al. 2001), they would get a 
shape factor of $\Gamma \simeq 0.65$ in an Einstein-de Sitter Universe, 
which is very far from the value we get here for the same cosmology, i.e. 
$\Gamma = 0.12$.  It is well known that the standard CDM model is ruled out 
in an Einstein-de Sitter cosmology, which is the reason for needing a 
non-standard CDM spectra in the $\Omega_{\rm M} \simeq 1$ case. This 
difference explains the apparently conflicting conclusions between V2003 and 
the present work. 
While the Einstein-de Sitter and concordance models produce distinct 
cluster number counts for a flux limited survey (Vauclair et al. 2003) or 
for the temperature distribution function evolution (Blanchard et al. 2000), 
once models are properly normalized at low redshift, the baryon mass function 
does not differentiate among flat cosmologies, because the inferred baryon 
masses are different, and this difference accidentally and roughly compensates 
for the effect of number evolution.
This kind of non-standard spectra with lower $\Gamma$ on cluster 
scales could have originated from different hypotheses on the dark matter 
content and/or from the existence of some unknown phase in the evolution of 
the Universe (like hot dark matter or quintessence) that could drive different 
power spectra for matter. The initial power spectrum could also be altered 
from a single power law by physics at the inflation period.
We are therefore lead to the final conclusion that present baryon cluster 
data can be described equally well by either a concordance or an 
Einstein-de Sitter 
(with a non-standard $\Gamma$ value on cluster scales) Universe 
implying that the baryon mass function is not as effective as the 
evolution of the temperature function. 
This removes the apparent discrepancies between the conclusions 
inferred from the $XMM-Newton~\Omega$-project (Vauclair et al. 2003, 
Sadat et al. 2005) and V2003.

\begin{acknowledgements}
      
SEN acknowledges M. Douspis for useful discussions and hospitality during 
a recent visit at LATT. He also thanks R. Pell\'o, L. Ferramacho, and 
M. Treguer for hospitality and the LENAC (Latin-american European 
Network for Astrophysics and Cosmology) project and CONICET for financial 
support. 
We are grateful to A. Vikhlinin for useful discussions and for providing us 
with an electronic version of the baryonic mass data. We also acknowledge 
the anonymous referee for several comments that helped to improve this paper.

\end{acknowledgements}

\end{document}